\documentclass[letterpaper,10pt,conference]{ieeeconf}

\IEEEoverridecommandlockouts
\usepackage{amsmath,graphicx}
\usepackage{amsfonts,amssymb}
\usepackage{booktabs}
\usepackage{multirow}
\usepackage{subfigure}
\usepackage{verbatim}
\usepackage{cite}

\def\L{{\cal L}}

\title{\LARGE \bf
Semi-Supervised Sound Event Detection with Conditional Mixup and Embedding-Level Contrastive Loss
}

\author{Nian Shao$^{1,2}$, Xian Li$^{2,3}$, Xiaofei Li$^{2,3*}$%
\thanks{$^{1}$Zhejiang University, Hangzhou, China.}%
\thanks{$^{2}$School of Engineering, Westlake University, Hangzhou, China}
\thanks{$^{3}$Westlake University \& Westlake Institute for Advanced Study, Hangzhou, China.}%
\thanks{$^{*}$Corresponding author. Email: lixiaofei@westlake.edu.cn}%
}

\begin{document}

\maketitle
\thispagestyle{empty}
\pagestyle{empty}

\begin{abstract}
Sound event detection (SED) is a core module for acoustic environmental analysis, yet its performance is often limited by scarce labeled data. Recent systems leverage large pretrained audio foundation models, but effective fine-tuning remains challenging because labeled data are limited while unlabeled data are abundant. A previous work, ATST-SED, addressed this problem with a pseudo-label based semi-supervised fine-tuning framework. In this work, we further improve the framework by adopting an embedding-level self-supervised contrastive loss inspired by ATST-Frame pretraining. This contrastive objective better exploits unlabeled data during fine-tuning. One challenge is that mixup serves different roles in the two objectives: pseudo-label learning uses composition mixup, while contrastive learning treats mixup as a perturbation. To resolve this mismatch, we propose conditional mixup, which combines composition mixup and perturbation mixup in one semi-supervised framework and defines the corresponding embedding-level contrastive losses. The resulting model achieves 0.645 PSDS$_1$ and 0.822 PSDS$_2$ on the DESED validation set, establishing a new state of the art.
\end{abstract}


\section{Introduction}
\label{sec:intro}
Sound conveys rich information about the surrounding environment, but recognizing sound events remains challenging for machines \cite{turpault2019sound}. Sound event detection (SED) aims to detect sound events and their temporal boundaries in polyphonic audio scenes. Since frame-level annotation is expensive, SED suffers from severe label scarcity. In practice, SED systems are trained with multiple annotation types. These include weakly labeled data, which provide clip-level event tags without temporal boundaries \cite{su2017weakly}; strongly labeled synthetic and real data, which provide time-aligned event annotations \cite{turpault2019sound,gemmeke2017audio}; and unlabeled data \cite{serizel2019trends}. This setup makes semi-supervised learning a natural choice for modern SED benchmarks. Prior work has improved semi-supervised SED with data augmentation, pseudo-label based learning, and stronger architectures for frame-level modeling \cite{koh2021sound,jiakai2018mean,zheng21skcrnn,nam22fdy}.

Recent SED systems also benefit from pretrained self-supervised audio models, including BEATs, AST, and ATST-Frame \cite{Chen2023BEATs,li2022atst,li2024self,gong21b_interspeech}. Trained on large-scale audio corpora, these models provide stronger representations than task-specific training alone. In particular, ATST-Frame is trained with a frame-level self-supervised contrastive objective and learns frame-level audio representations that transfer well to temporally localized downstream tasks such as SED \cite{li2024self}. In prior ATST-SED work \cite{shao2024atst_sed}, ATST-Frame was introduced into a CRNN-based detector, and a two-stage pseudo-label based semi-supervised framework was proposed for fine-tuning ATST-Frame for SED, providing a strong baseline.

However, the semi-supervised objectives commonly used in SED are dominated by consistency regularization \cite{koh2021sound,shao22rct,guan2024sod}. These objectives are effective for lightweight SED models, but can be much less effective for SED systems built on large pretrained self-supervised encoders \cite{shao2022atst, liu2023, Li2023panns}. A likely reason is that consistency regularization is too weak compared with the much stronger pretraining tasks. By contrast, self-supervised contrastive learning is a more challenging representation-learning objective and has been shown to improve the linear separability and transferability of learned representations during pretraining \cite{li2024self,chen2021exploring}. Among existing self-supervised contrastive frameworks, ATST-Frame is particularly suitable for SED because it is trained with a frame-level objective that matches the frame-level nature of SED. We therefore introduce its embedding-level self-supervised contrastive objective into semi-supervised fine-tuning, so that unlabeled data can be better leveraged than pseudo-label consistency alone.

Pseudo-label learning and contrastive learning, however, benefit from different uses of mixup. In pseudo-label learning, mixup follows a composition view: the mixed sample should preserve the content of both source samples. In contrastive learning, mixup follows a perturbation view: the newly added sample should only perturb the input sample \cite{niizumi2021byol-a,li2023ast,li2024self}. Both strategies are effective in their own settings. To exploit both of them within one framework, we propose conditional mixup, which combines composition mixup and perturbation mixup according to the interpolation coefficient. Based on the two mixup cases, we further define the corresponding pseudo-label losses and embedding-level contrastive losses, so that decision-level pseudo-label supervision and embedding-level contrastive supervision can complement each other during fine-tuning. On the DESED validation set, ATST-SEDv2 achieves 0.645 PSDS$_1$ and 0.822 PSDS$_2$ with cSEBB post-processing, establishing a new state of the art. The main contributions are as follows. First, we propose conditional mixup to combine composition mixup and perturbation mixup in one framework. Second, we design the corresponding pseudo-label and embedding-level contrastive losses, which provide an effective semi-supervised objective for both lightweight and pretraining-based SED models. Third, we achieve new state-of-the-art performance on the DESED validation set, reaching 0.645 PSDS$_1$ and 0.822 PSDS$_2$.

\section{Method}

Polyphonic SED is a multi-class detection task in which a model detects the onset and offset of multiple sound events in an audio clip. This work focuses on the DESED dataset \cite{turpault2019sound}, which provides strongly labeled, weakly labeled, and unlabeled training data. To make use of both supervised and unsupervised data, the SED model is trained in a semi-supervised manner. In this work, we aim to strengthen semi-supervised fine-tuning for ATST-SED by adopting an embedding-level self-supervised contrastive objective inspired by ATST-Frame pretraining \cite{li2024self}. In this section, we first review the pseudo-label based semi-supervised fine-tuning baseline. We then describe the mixup strategies used in the baseline system and in contrastive learning, and introduce conditional mixup to unify them. Next, we describe the corresponding embedding-level self-supervised losses and summarize the overall training objective.

\subsection{Pseudo-Label Based Two-Stage Fine-Tuning Baseline}

\begin{figure}
    \centering
    \includegraphics[width=0.8\linewidth]{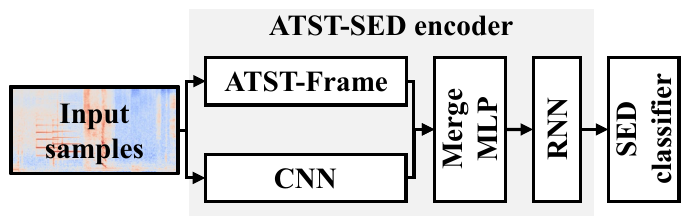}
    \caption{Model architecture of the ATST-SED model, which consists of an ATST-SED encoder and a SED classifier.}
    \vspace{-0.8em}
    \label{fig: atst_arch}
\end{figure}

We use ATST-SED as the baseline model \cite{shao2024atst_sed}. As shown in Fig.~\ref{fig: atst_arch}, the model consists of an ATST-SED encoder and a nonlinear SED classifier \cite{jiakai2018mean}. The encoder consists of a convolutional recurrent neural network (CRNN) \cite{jiakai2018mean} and a pretrained ATST-Frame model \cite{li2024self}. The pretrained ATST-Frame model serves as an additional feature extractor alongside the 7-layer convolutional neural network (CNN) feature extractor. The features from the two branches are temporally aligned and fused, after which a recurrent neural network (RNN) performs contextual modeling before classification.

The baseline fine-tuning framework is divided into two stages. Both stages are trained in a semi-supervised manner. In stage 1, the ATST-Frame model is kept frozen, and only the CRNN and SED classifier are optimized. The supervised binary cross-entropy (BCE) loss ($\L_{\text{BCE}}$) is applied to the labeled data. For unlabeled data, the Mean Teacher (MT) loss ($\L_{\text{MT}}$) \cite{tarvainen2017mean} is used. Specifically, an exponential moving average (EMA) teacher generates pseudo labels for the unlabeled data, and the MT loss minimizes the mean-square error between the student predictions and these pseudo labels. The total loss of stage 1 is
\begin{align}
    \L_\text{stage1}=\L_\text{BCE} + r_\text{MT}\L_\text{MT},
\end{align}
where $r_\text{MT}$ is the ramp-up weight of $\L_\text{MT}$.
In stage 2, all parameters are fine-tuned together. Since the amount of labeled data in DESED is still limited for tuning a large pretrained encoder, the baseline further applies mixup and frequency warping (FW) \cite{li2024self} to 50\% of the training data to increase data diversity, and assigns larger weights to the unsupervised losses to make fuller use of the unlabeled data. Besides BCE and MT, interpolation consistency training (ICT) loss ($\L_\text{ICT}$) \cite{verma2019interpolation} is used together with mixup for semi-supervised learning. When mixup is applied, the ICT loss replaces the MT loss for the mixed samples. The total loss of stage 2 is
\begin{align}
    \L_\text{stage2}=\L_\text{BCE} + r_\text{MT}\L_\text{MT} + r_\text{ICT}\L_\text{ICT},
\end{align}
where $r_\text{ICT}$ is the ramp-up weight of $\L_\text{ICT}$. In this work, we keep the stage-1 training unchanged and only redesign the stage-2 fine-tuning strategy.

\begin{figure*}[!t]
    \centering
    \includegraphics[width=0.64\linewidth]{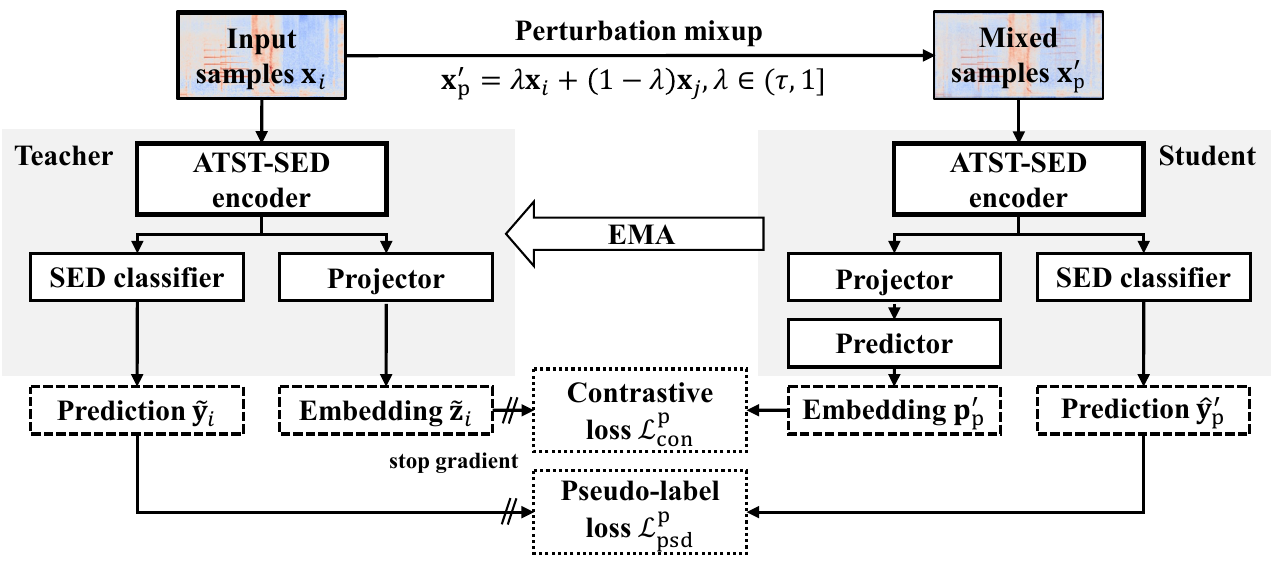}
    \caption{Flowchart of the proposed pseudo-label and contrastive losses, when using perturbation mixup for data augmentation.}
    \vspace{-0.8em}
    \label{fig: flowchart}
\end{figure*}

\subsection{Conditional mixup}
The mixup \cite{zhang2017mixup} used in the baseline model combines two samples and treats the sound events from both audio clips as co-occurring in the mixed sample. Formally, for an input sample $(\mathbf{x}_i, \mathbf{y}_i)$ and a randomly selected auxiliary sample $(\mathbf{x}_j, \mathbf{y}_j)$, where $\mathbf{x} \in \mathbb{R}^{T\times D}$ and $\mathbf{y} \in \mathbb{R}^{T\times C}$, $T$ and $D$ denote the time and feature dimensions, and $C$ is the number of classes, mixup generates a new sample $(\mathbf{x}', \mathbf{y}')$ as $\mathbf{x}' = \lambda \mathbf{x}_i + (1-\lambda)\mathbf{x}_j$, and
$\mathbf{y}' = \lambda \mathbf{y}_i + (1-\lambda) \mathbf{y}_j$,
where $\lambda \in [0,1]$ is typically drawn from a Beta distribution. 
In semi-supervised learning, $\tilde{\mathbf{y}}_i$ and $\tilde{\mathbf{y}}_j$ denote the teacher predictions on $\mathbf{x}_i$ and $\mathbf{x}_j$, respectively. The mixed pseudo-label is $\lambda \tilde{\mathbf{y}}_i + (1-\lambda)\tilde{\mathbf{y}}_j$. Denote the student prediction on $\mathbf{x}'$ as $\hat{\mathbf{y}}'$. The corresponding pseudo-label loss is
\begin{align}
\L_{\text{ICT}} = \left\|\hat{\mathbf{y}}' - (\lambda \tilde{\mathbf{y}}_i + (1-\lambda)\tilde{\mathbf{y}}_j)\right\|_2^2,
\end{align}
which is the ICT loss \cite{verma2019interpolation}. Since this mixup preserves the content of both source clips, we refer to it as \textbf{composition mixup}.

Audio contrastive self-supervised learning also adopts mixup \cite{niizumi2021byol-a, li2023ast, li2024self}, but treats the auxiliary sample as a perturbation to the input sample. In this context, $\lambda$ is typically drawn from a restricted uniform distribution (e.g., in \cite{niizumi2021byol-a}, $\lambda \sim \textbf{U}(0.6, 1)$), which ensures that the contribution of the auxiliary sample remains smaller than that of the input sample. We refer to this use of mixup as \textbf{perturbation mixup}. When using it for semi-supervised learning, the mixed sample $\mathbf{x}'$ is constructed in the same way, while only the target of the input sample is preserved, i.e., $\mathbf{y}' = \mathbf{y}_i$ and $\tilde{\mathbf{y}}_i$ is used as the pseudo label in pseudo-label loss.

To unify the two mixup methods, we propose \textbf{conditional mixup}. In audio mixup, the interpolation coefficient $\lambda$ can be interpreted as the relative energy ratio between the input signal and the auxiliary signal, which is proportional to $\frac{\lambda}{1-\lambda}$. When $\lambda$ is close to 0.5, the two signals have similar energy, so events from both signals are treated as co-occurring in the mixture. When $\lambda$ is large, the input signal dominates the mixture, and the auxiliary signal is treated as a perturbation. We therefore use $\lambda \in [0.5, 1]$ together with a threshold $\tau$ to determine the mixup mode. When $0.5 \leq \lambda \leq \tau$, the mixed sample follows the composition view. In this case, we use label combination rather than interpolation to better represent event co-occurrence: the supervised target is the logical OR $\mathbf{y}_i \lor \mathbf{y}_j$, while the semi-supervised target is the sum $\tilde{\mathbf{y}}_i + \tilde{\mathbf{y}}_j$, clamped by maximum of 1. Formally, for $0.5 \leq \lambda_{\text{c}} \leq \tau$, we define

\begin{align}
\mathbf{x}'_{\text{c}} &= \lambda_{\text{c}}\mathbf{x}_i + (1-\lambda_{\text{c}})\mathbf{x}_j,\\
\mathbf{y}'_{\text{c}} &= \mathbf{y}_i \lor \mathbf{y}_j.
\end{align}
The corresponding pseudo-label loss $\L_{\text{c-psd}}$ is
\begin{align}
\L_{\text{c-psd}} = \left\|\hat{\mathbf{y}}'_{\text{c}} - \text{min}(\tilde{\mathbf{y}}_i + \tilde{\mathbf{y}}_j, \mathbf{1})\right\|_2^2,
\end{align}
When $\tau < \lambda \leq 1$, the mixed sample follows perturbation mixup, and only the target of the input sample is preserved. For $\tau < \lambda_{\text{p}} \leq 1$, we define
\begin{align}
\mathbf{x}'_{\text{p}} &= \lambda_{\text{p}}\mathbf{x}_i + (1-\lambda_{\text{p}})\mathbf{x}_j,\\
\mathbf{y}'_{\text{p}} &= \mathbf{y}_i.
\end{align}
The corresponding pseudo-label loss $\L_{\text{p-psd}}$ is
\begin{align}
\L_{\text{p-psd}} = \left\|\hat{\mathbf{y}}'_{\text{p}} - \tilde{\mathbf{y}}_i\right\|_2^2,
\end{align}
The unified pseudo-label objective for conditional mixup is
\begin{align}
\L_{\text{cp-psd}} = \frac{1}{2}\left(\L_{\text{c-psd}} + \L_{\text{p-psd}}\right).
\end{align}

\subsection{Embedding-Level Semi-Supervised Contrastive Losses}

We now introduce embedding-level semi-supervised contrastive losses. Following \cite{li2024self}, contrastive learning forms positive pairs by applying data augmentation to the same sample and aligning the resulting embeddings. In our framework, conditional mixup produces two augmented views for each input, corresponding to the composition and perturbation cases. We therefore define two contrastive losses accordingly. As shown in Fig.~\ref{fig: flowchart}, a projector and a predictor are attached after the ATST-SED encoder for contrastive learning. The contrastive loss is computed between the student predictor output and the teacher projector output, which helps prevent collapse in contrastive learning \cite{chen2021exploring}. The extracted embeddings are $\ell_2$-normalized along the hidden dimension.

\textbf{Perturbation case} For the positive pair $(\mathbf{x}_i, \mathbf{x}'_\text{p})$ created by perturbation mixup, the auxiliary sample is treated only as a perturbation to the input sample. Therefore, the contrastive objective is to maximize the similarity between the mixed sample and the input sample. Let $\tilde{\mathbf{z}}_i \in \mathbb{R}^{T\times H}$ denote the teacher embedding of $\mathbf{x}_i$, and let $\mathbf{p}'_\text{p} \in \mathbb{R}^{T\times H}$ denote the student embedding of the mixed sample $\mathbf{x}'_\text{p}$. The frame-level contrastive loss is defined as
\begin{align}
\L_\text{con}^\text{p}(\mathbf{x}_i, \mathbf{x}'_\text{p}) = \frac{1}{T}\sum_{t=1}^{T} \|\tilde{\mathbf{z}}_i(t, :) - \mathbf{p}'_\text{p}(t, :)\|_2^2,
\end{align}
where $(t, :)$ denotes the embedding at the $t$-th frame. As in \cite{li2024self}, we also use the reversed positive pair by feeding $\mathbf{x}'_\text{p}$ to the teacher and $\mathbf{x}_i$ to the student, and denote the resulting loss by $\L^{\prime\text{p}}_\text{con}$. The total perturbation contrastive loss is
\begin{align}
\L_\text{p-con}=\frac{1}{2}(\L_\text{con}^\text{p} + \L^{\prime\text{p}}_\text{con}).
\end{align}

\textbf{Composition case} For the positive pair $(\mathbf{x}_i, \mathbf{x}'_\text{c})$ created by composition mixup, both the input sample and the auxiliary sample contribute to the mixed content. Therefore, the contrastive objective is to maximize the similarity between the mixed sample and both source samples. Let $\tilde{\mathbf{z}}_i, \tilde{\mathbf{z}}_j \in \mathbb{R}^{T\times H}$ denote the teacher embeddings of $\mathbf{x}_i$ and $\mathbf{x}_j$, and let $\mathbf{p}'_\text{c} \in \mathbb{R}^{T\times H}$ denote the student embedding of the mixed sample $\mathbf{x}'_\text{c}$. We use the average of the two teacher embeddings as the contrastive target. The contrastive loss is defined as
\begin{align}
\L_\text{con}^{\text{c}}(\mathbf{x}_i, \mathbf{x}'_\text{c}) = \frac{1}{T}\sum_{t=1}^{T} \left\|\frac{1}{2}\left(\tilde{\mathbf{z}}_i(t, :) + \tilde{\mathbf{z}}_j(t, :)\right) - \mathbf{p}'_\text{c}(t, :)\right\|_2^2.
\end{align}
Similarly, we also use the reversed positive pair and denote the corresponding loss by $\L^{\prime\text{c}}_\text{con}$. The total composition contrastive loss is
\begin{align}
\L_\text{c-con}=\frac{1}{2}(\L_\text{con}^{\text{c}} + \L^{\prime\text{c}}_\text{con}).
\end{align}
The total contrastive loss used in this work is
\begin{align}
\L_\text{con}=\frac{1}{2}(\L_\text{c-con} + \L_\text{p-con}).
\end{align}

\subsection{Overall training objective}
The proposed method is used for stage-2 fine-tuning of ATST-SED. The overall training objective is composed of three parts: the supervised BCE loss, the pseudo-label based loss, and the contrastive loss. We use the proposed conditional mixup for data augmentation. Frequency warping \cite{li2024self} is also applied to the mixed samples to further increase the difficulty of the semi-supervised task. The unmixed input samples remain unaugmented so that the teacher model can generate reliable pseudo labels, while the original Mean Teacher loss remains unchanged. The overall training objective is
\begin{align}
\L_{\text{total}} = \L_\text{BCE} + r_\text{MT}\L_\text{MT} + r_\text{cp-psd}\L_{\text{cp-psd}} + r_\text{con}\L_\text{con},
\end{align}
where $r_\text{cp-psd}$ and $r_\text{con}$ are the weights for the pseudo-label loss and contrastive loss, respectively.

\section{Experimental Results}
\label{sec:exp}
\subsection{Experiment setup}
Code is available on our website\footnote{https://github.com/Audio-WestlakeU/ATST-SED}. All experiments are conducted on DESED \cite{turpault2019sound}. The training set comprises 1,578 weakly labeled clips, 10,000 synthesized strongly labeled clips, 14,412 unlabeled clips, and 3,470 additional real strongly labeled clips from AudioSet \cite{gemmeke2017audio}. The evaluation is performed on the DESED validation set with 1,168 strongly labeled audio clips. For preprocessing, all recordings are resampled to 16 kHz and truncated or padded to 10 seconds. For ATST-SED, the CNN branch uses 128-dimensional LogMel features with a frame length of 128 ms and a hop length of 16 ms. The ATST-Frame branch keeps its original preprocessing pipeline.

When applying the proposed method, we add a projector and a predictor only during training for embedding-level contrastive regularization, while keeping the inference architecture unchanged. Both modules share the same architecture: a linear layer, batch normalization, a ReLU, and a final linear layer. The batch sizes for real strongly labeled, synthetic strongly labeled, weakly labeled, and unlabeled data are 24, 24, 48, and 48, respectively. We optimize the models with Adam \cite{kingma2015adam}. For ATST-SED stage-2 fine-tuning, the learning-rate setup is the same as in \cite{shao2024atst_sed}. A layer-wise decaying learning rate is used for the ATST blocks (initial rate 2e-4, decaying by 0.5 per layer), and a uniform initial rate of 2e-4 is used for the CNN, projector, and predictor. The RNN and classifier use a higher initial rate of 2e-3. Fine-tuning lasts for 20,000 steps. All learning rates follow an exponential ramp-up from 0 to their initial values during the first 290 steps. They then cosine-decay to $\frac{\alpha}{10}$. The hyperparameters are empirically set as follows: $r_{\text{MT}}=70$, $r_{\text{cp-psd}}=17.5$, $r_{\text{con}}=3$, and the mixup-mode threshold $\tau=0.55$.

Post-processing applies median filtering and change-detection-based sound event bounding boxes (cSEBB) \cite{ebbers24sebb} to transform soft model predictions into binary decisions. Models are evaluated every 10 epochs on the DESED validation set using PSDS$_1$ and PSDS$_2$ \cite{ebbers2022psds}. PSDS$_1$ emphasizes temporal stability, and PSDS$_2$ emphasizes audio tagging accuracy, where higher values indicate better performance. Unless otherwise specified, all reported results use median-filter post-processing.

\subsection{Comparison with the baseline model}
\begin{figure}
    \includegraphics[width=\linewidth]{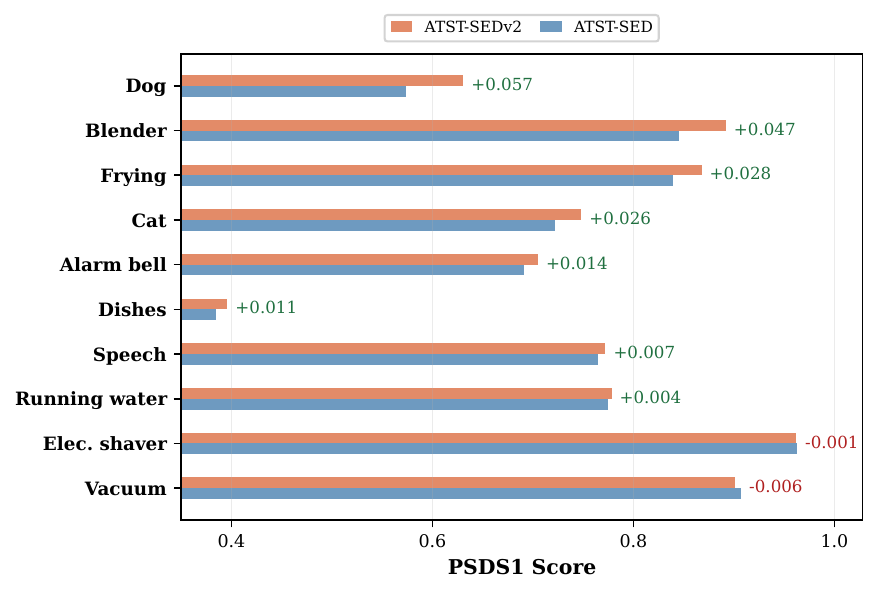}
    \vspace{-1.5em}
    \caption{Comparison between ATST-SED \cite{shao2024atst_sed} and the proposed ATST-SEDv2 on per-class PSDS1 scores.}
    \label{fig: per_class_score}
\end{figure}

We first apply the proposed method to the baseline models, and the results are shown in Table~\ref{tab:prev_compare}. We also train a CRNN model to show that the proposed method is effective on lightweight SED models. Its training follows the DCASE 2023 baseline setup. When applying our method to CRNN, we use $r_{\text{cp-psd}}=0.125$, $r_{\text{con}}=0.1$, and $\tau=0.55$. For clarity, we denote ``ATST-SED + proposed method'' as ATST-SEDv2. Our approach obtains consistent improvements across both architectures. 
For PSDS$_1$, CRNN improves from 0.384 to 0.403, and ATST-SED improves from 0.583 to 0.607. These gains are non-trivial for PSDS$_1$, because this metric strongly penalizes instability across classes~\cite{ebbers2022psds}. For PSDS$_2$, CRNN improves from 0.628 to 0.661, and ATST-SED improves from 0.810 to 0.817, indicating improved tagging capacity while maintaining temporal detection quality. The stronger gain on CRNN suggests that the method is especially helpful when the backbone capacity is limited, while the gain on ATST-SED shows that the method is still effective on a strong pretrained baseline. 
Per-class analysis in Fig.~\ref{fig: per_class_score} compares our system with ATST-SED, showing that the improvement is distributed across classes rather than concentrated in only a few easy classes. The method particularly benefits challenging categories such as ``dog'', ``cat'', and ``dishes''. These events are short and have high acoustic variability, so they are difficult for SED systems~\cite{serizel2019trends}. 
Overall, the proposed approach effectively enhances semi-supervised learning across both lightweight and pretraining-based SED models.

\begin{table}[t]
\caption{Comparison with the previous baselines.}
\vspace{-0.8em}
\label{tab:prev_compare}
\centering
\begin{tabular}{lcc}
\toprule
\textbf{Method} & \textbf{PSDS$_1$} & \textbf{PSDS$_2$} \\
\midrule
CRNN                                & 0.384 & 0.628 \\
CRNN + proposed method              & \textbf{0.403} & \textbf{0.661} \\
\hline
ATST-SED \cite{shao2024atst_sed}    & 0.583 & 0.810 \\
ATST-SED + proposed method (ATST-SEDv2)        & \textbf{0.607} & \textbf{0.817} \\
\bottomrule
\end{tabular}
\vspace{-0.8em}
\end{table}

\subsection{Ablation Study}

\begin{table}[t]
\caption{Ablation study of ATST-SEDv2 components.}
\vspace{-0.8em}
\label{tab:ablation_v2}
\centering
\begin{tabular}{lcc}
\toprule
\textbf{Method} & \textbf{PSDS$_1$} & \textbf{PSDS$_2$} \\
\midrule
Stage 1 frozen training                     & 0.529  & 0.778 \\
Stage 2                                     &       & \\
\quad w. $\L_\text{con}$                    & 0.560 & 0.811 \\
\quad w. $\L_\text{MT}+\L_\text{cp-psd}$    & 0.579 & 0.800 \\
\quad w. $\L_\text{MT} + \L_\text{p-psd} + \L_\text{p-con}$ & 0.595 & 0.810 \\
\quad w. $\L_\text{MT} + \L_\text{c-psd} + \L_\text{c-con}$ & 0.601 & 0.817 \\
\quad w. $\L_\text{MT} + \L_\text{cp-psd} + \L_\text{con}$ (ATST-SEDv2)                       & 0.607 & 0.817 \\
\bottomrule
\end{tabular}
\vspace{-0.8em}
\end{table}

Table~\ref{tab:ablation_v2} reports the ablation results. Each row corresponds to an independent stage-2 fine-tuning setting, and the stage-1 result is listed for reference. Using only the contrastive loss ($\L_\text{con}$) improves performance over stage 1, which confirms that the contrastive objective is useful in semi-supervised fine-tuning. Pseudo-label losses ($\L_\text{MT}+\L_\text{cp-psd}$) remain essential, and using them alone yields performance close to ATST-SED~\cite{shao2024atst_sed}. However, the gain from contrastive loss is clearly smaller than that from pseudo-label losses. A likely reason is that the contrastive objective and the SED objective do not necessarily share the same optimum. When contrastive loss is used alone, the available supervised data may be too limited to regularize the network toward the SED optimum. Therefore, contrastive loss should be used together with pseudo-label losses to better leverage unlabeled data during fine-tuning.

For conditional mixup, we further ablate the composition and perturbation branches separately while keeping the Mean Teacher loss. Both the composition branch ($\L_\text{c-psd} + \L_\text{c-con}$) and the perturbation branch ($\L_\text{p-psd} + \L_\text{p-con}$) are effective when used separately. In particular, the composition-only setting reaches 0.601/0.817, which is clearly better than the original ATST-SED result of 0.583/0.810. This result shows that the composition branch, together with the introduced contrastive loss $\L_\text{c-con}$, is effective. Similarly, the perturbation-only setting also yields strong results, validating perturbation mixup as an effective augmentation for semi-supervised learning. Overall, each proposed module improves the stage-1 model when used individually, and the modules are compatible. Combining all components yields the best performance. 

\subsection{Comparison with SOTA SED systems}

\begin{table}[t]
\renewcommand\arraystretch{1.2}
\caption{Comparison with SOTA SED models (MF: median filter). Methods marked with \textsuperscript{\dag} use the scores reported in their original papers, and PSDS$_2$ is not provided for them.}
\vspace{-0.8em}
\label{tab: sota_results}
\centering
\begin{tabular}{l|cc|cc}
\toprule
\multirow{2}{*}{\textbf{Method}} & \multicolumn{2}{|c}{\textbf{PSDS$_1$}} & \multicolumn{2}{|c}{\textbf{PSDS$_2$}}  \\
\cline{2-5}
& cSEBB & MF & cSEBB & MF \\
\midrule
ATST-SED \cite{shao2024atst_sed}    & 0.618 & 0.583 & 0.813 & 0.810 \\
MAT-SED\textsuperscript{\dag} \cite{cai24matsed}    & 0.602 & 0.587 &  - & - \\
PMAM$_\text{iter2}$\textsuperscript{\dag} \cite{cai2025prototype}   & 0.625 & 0.597 & - & - \\
ATST-SEDv2 (ours)   & \textbf{0.645} & \textbf{0.607} & \textbf{0.822} & \textbf{0.817} \\
\bottomrule
\end{tabular}
\vspace{-0.8em}
\end{table}

Table~\ref{tab: sota_results} benchmarks ATST-SEDv2 against competing SOTA methods on the DESED validation set. To ensure a fair comparison, we apply cSEBB to both our model and the ATST-SED baseline. cSEBB uses adaptive thresholds, rather than a single threshold, to detect event boundaries from soft predictions. This allows both low-confidence and high-confidence events to be detected and generally yields more accurate results.

We compare with two recent methods: MAT-SED~\cite{cai24matsed} and PMAM$_{\text{iter2}}$~\cite{cai2025prototype}. Both methods apply self-supervised learning to AST-based \cite{gong21b_interspeech} SED systems. They first pretrain with self-supervised tasks and then perform semi-supervised fine-tuning with pseudo-label losses. PMAM further applies iterative updates with prototypical modeling and achieves strong PSDS$_1$ performance. Compared with these methods, ATST-SEDv2 uses a simpler strategy by jointly applying contrastive and pseudo-label losses during fine-tuning. Fine-tuning can be completed in a single run with lower training complexity. Even with this simpler design, the method leverages unlabeled data effectively and achieves better PSDS$_1$ performance under both cSEBB and median-filter post-processing.

Overall, ATST-SEDv2 establishes new state-of-the-art results on the DESED validation set, reaching 0.645 PSDS$_1$ and 0.822 PSDS$_2$.

\section{Conclusion}
In this paper, we extend ATST-SED by introducing an embedding-level self-supervised contrastive objective into semi-supervised fine-tuning. Since pseudo-label learning and contrastive learning favor different mixup semantics, we propose conditional mixup to combine composition mixup and perturbation mixup in one framework, together with the corresponding pseudo-label and embedding-level contrastive losses. In this way, decision-level pseudo-label supervision and embedding-level contrastive supervision complement each other during fine-tuning. This design improves the use of unlabeled data while keeping the inference-time model unchanged. On the DESED validation set, the proposed method achieves state-of-the-art performance, reaching 0.645 PSDS$_1$ and 0.822 PSDS$_2$.

\clearpage
\bibliographystyle{IEEEtran}
\bibliography{refs}
\end{document}